\documentclass{IEEEtran}

\ifCLASSINFOpdf
  \usepackage[pdftex]{graphicx}
  \graphicspath{{figures/}}
  \DeclareGraphicsExtensions{.pdf,.jpeg,.png, .eps}
\else
\fi

\usepackage[utf8]{inputenc}
\usepackage[T1]{fontenc}
\usepackage[nolist]{acronym}
\usepackage[colorlinks=true,allcolors=blue]{hyperref}
\usepackage{url}
\usepackage{pifont}
\usepackage{multirow}
\usepackage{siunitx}
\usepackage{array,booktabs}

\usepackage[usenames,dvipsnames,svgnames,table]{xcolor}
\usepackage[utf8]{inputenc}
\usepackage{tikz}
\usetikzlibrary{snakes,arrows,shapes}
\usepackage{amsmath}
\usepackage{color,soul}

\definecolor{gray}{gray}{0.9}
\definecolor{white}{gray}{1.0}
\newcolumntype{g}{>{\columncolor{gray}}l}
\newcolumntype{x}{>{\columncolor{gray}}p{1cm}}
\newcolumntype{y}{>{\columncolor{gray}}p{4cm}}
\newcolumntype{z}{>{\columncolor{gray}}p{2cm}}
\begin{document}

\begin{acronym}
  \acro{NETCONF}{Network Configuration}
  \acro{IETF}{Internet Engineering Task Force}
  \acro{LMAP}{Large-Scale Measurement of Broadband Performance}
  \acro{MA}{Measurement Agent}
  \acro{CPE}{Customer Premises Equipment}
  \acro{NETMOD}{NETCONF Data Modeling Language}
  \acro{TLS}{Transport Layer Security}
  \acro{SSH}{Secure Shell}
  \acro{NMS}{Network Management System}
  \acro{NAT}{Network Address Translation}
  \acro{TCP}{Transmission Control Protocol}
  \acro{NMS}{Network Management System}
  \acro{CH}{Call Home}
  \acro{MAs}{Measurement Agents}
  \acro{DDoS}{Distributed Denial-of-Service}
  \acro{LMAP}{Large-Scale Measurement of Broadband Performance}
\end{acronym}

\title{Standards Support in NETCONF Tools}
\title{Standards Development in Open NETCONF Tools}
\title{Towards Managing Home Routers using NETCONF}
\title{Implementing NETCONF over TLS and NETCONF Call Home to Manage Home Routers}
\title{Managing Home Routers with NETCONF over TLS and NETCONF Call Home}

 %affiliations

\author{\IEEEauthorblockN{Vaibhav Bajpai\IEEEauthorrefmark{1}, Radek
Krejci\IEEEauthorrefmark{2} and Leonidas Poulopoulos\IEEEauthorrefmark{3}}

\IEEEauthorblockA{\IEEEauthorrefmark{1}Technical University of Munich \\
\texttt{bajpaiv@in.tum.de}}

\IEEEauthorblockA{\IEEEauthorrefmark{2}CESNET \\
\texttt{rkrejci@cesnet.cz}}

\IEEEauthorblockA{\IEEEauthorrefmark{3}Verisign \\
\texttt{lpoulopoulos@verisign.com}}}

\maketitle

\begin{abstract}

  The \ac{NETCONF} protocol and the associated YANG data modeling language are
  the foundations of contemporary network management frameworks evolving
  within the \ac{IETF}. \texttt{netopeer} (a NETCONF server) and
  \texttt{ncclient} (a NETCONF client) are popular open-source projects that
  support the latest NETCONF v1.1 protocol using the mandatory \ac{SSH}
  transport. We recently implemented and integrated NETCONF over \ac{TLS}
  transport and NETCONF \ac{CH} mechanisms using reverse \ac{TLS} and \ac{SSH}
  in both projects. The \ac{CH} mechanism allows a managed device behind a
  \ac{NAT} running a \ac{NETCONF} server (\texttt{netopeer}) to successfully
  establish a \ac{NETCONF} session with a \ac{NMS} running a \ac{NETCONF}
  client (\texttt{ncclient}). In this article, we describe how these standards
  allow home routers and \ac{NAT} boxes (in particular) to be managed using
  these latest additions to the \ac{NETCONF} protocol.

  %alongwith with perspectives on future standardization efforts currently
  %being pursued within the \ac{IETF}.

  %Interoperability testing is an important part of the lifecycle of a protocol.
  %Interoperability tests not only help to reduce implementation bugs, but they
  %also often improve the accuracy of a protocol specification. In a pursuit to
  %improve \ac{NETCONF} interoperability and foster it for community-based
  %education, we have developed and deployed a \ac{NETCONF} Interoperability Lab.
  %The lab provides a platform to execute test-cases against an online catalogue
  %of \ac{NETCONF} server and client implementations. The NETCONF
  %Interoperability lab is available online at:
  %\texttt{\url{http://www.interop-lab.net.}}

\end{abstract}

%\begin{IEEEkeywords} NETCONF, YANG, IETF, TLS, SSH, NAT, Call Home, Home router\end{IEEEkeywords}

\IEEEpeerreviewmaketitle

% sections
%----------------------------------------------------------------------
\section{Introduction} \label{sec:introduction}

%There is a growing need for tools and related specifications to evaluate the
%interoperability and compliance of implementations of standardized network
%management protocols. Interoperability problems are typically caused by a
%number of different factors: a) Errors and ambiguities in standards. b) Human
%programming errors. c) Different interpretations of the standard. d) Different
%choice of options allowed by the standard.  As such, it is not uncommon that
%interoperability tests uncover ambiguous parts of a specification and parts
%that are under-specified. This is why, within the \ac{IETF}, interoperability
%reports are generally needed in order to advance specifications on the
%standards-track \cite{rfc5657:2009}.

%The availability of open labs that can be used for interoperability testing
%will help alleviate such challenges. These labs ease validation of reference
%implementations and foster their acceptance in the market place. The fact that
%major components of a testing lab are made publicly accessible (test
%specifications, tools, execution environments) will not only directly benefit
%the developers, but also potential users by providing a common testing ground
%within the lab. In this pursuit, we have developed and deployed an open lab
%that can be used for \ac{NETCONF} interoperability testing.

The \ac{NETCONF} protocol (RFC 6241) and the associated YANG data modeling
language (RFC 6020) are the foundations of a new network management frameworks
evolving within the \ac{IETF}.  \ac{NETCONF}~\cite{jschonwalder:commag:2010}
is a network management protocol that provides a secure mechanism to install,
manipulate and delete the configuration of network devices.  \ac{NETCONF} was
standardized by the \ac{IETF} in RFC 4741 and revised in RFC 6241.
\ac{NETCONF} is also tightly connected with another \ac{IETF} activity -- YANG
modeling language (RFC 6020, RFC 7950). YANG data models are used to define
not only the data accessible via \ac{NETCONF}, but also \ac{NETCONF}
operations, notifications and various extensions.

There has been work on \ac{NETCONF} interoperability testing in the
past~\cite{hmtran:aims:2009}. Meanwhile, implementations have matured
significantly and the core \ac{NETCONF} specifications have been revised in
order to remove ambiguities and to address problems encountered with the first
revision. For instance, Table~\ref{tab:capabilities} enlists basic
\ac{NETCONF} capabilities and protocol extensions supported by popular
open-source and commercial server implementations. The standardized YANG data
models supported by each server implementation are enlisted in
Table~\ref{tab:datamodels}.

%\hl{$--$ R's note: OpenYuma is not well supported/maintained, is it still
%worthwile to keep it in the tables?}

%\hl{$--$ Vaibhav, are you able to update the table for the current Yuma and
%ConfD versions? There should be probably some other features like support for
%YANG 1.1.}

\begin{table}[t!]

  \caption{\ac{NETCONF} capabilities support}
  \label{tab:capabilities}

	\begin{center}
		\begin{tabular}{lcccc}

    \multirow{2}{*}{Capability}        & ConfD           & YumaPro           & \texttt{Netopeer2} \\
                                       & \texttt{(v5.0)} & \texttt{(v13.04)} & \texttt{(v0.3)} \\

		\toprule

    \texttt{:base:1.0}                 & \ding{52}       & \ding{52}         & \ding{52}       \\
    \texttt{:base:1.1}                 & \ding{52}       & \ding{52}         & \ding{52}       \\
    \texttt{:writable-running:1.0}     & \ding{52}       & \ding{54}         & \ding{52}       \\
    \texttt{:candidate:1.0}            & \ding{52}       & \ding{52}         & \ding{52}       \\
    \texttt{:rollback-on-error:1.0}    & \ding{52}       & \ding{52}         & \ding{52}       \\
    \texttt{:startup:1.0}              & \ding{54}       & \ding{54}         & \ding{52}       \\
    \texttt{:url:1.0}                  & \ding{52}       & \ding{52}         & \ding{54}       \\
    \texttt{:xpath:1.0}                & \ding{52}       & \ding{52}         & \ding{52}       \\
    \texttt{:confirmed-commit:1.0}     & \ding{52}       & \ding{52}         & \ding{54}       \\
    \texttt{:confirmed-commit:1.1}     & \ding{52}       & \ding{52}         & \ding{54}       \\
    \texttt{:validate:1.0}             & \ding{52}       & \ding{52}         & \ding{52}       \\
    \texttt{:validate:1.1}             & \ding{52}       & \ding{52}         & \ding{52}       \\
    \texttt{:notification:1.0}         & \ding{54}       & \ding{52}         & \ding{52}       \\
    \texttt{:interleave:1.0}           & \ding{54}       & \ding{52}         & \ding{52}       \\
    \texttt{:partial-lock:1.0}         & \ding{54}       & \ding{52}         & \ding{54}       \\
    \texttt{:with-defaults:1.0}        & \ding{52}       & \ding{52}         & \ding{52}       \\

    \bottomrule
		\end{tabular}
	\end{center}
\end{table}

\begin{table}[t!]
  \caption{YANG data models support}
  \label{tab:datamodels}
	\begin{center}
		\begin{tabular}{lcccc}
    \multirow{2}{*}{Data Models}           & ConfD             & YumaPro            & \texttt{Netopeer2}   \\
                                           & \texttt{(v5.0)}   & \texttt{(v13.04)}  & \texttt{(v0.3)}   \\
		\toprule

    \texttt{ietf-inet-types}               & \ding{52}         & \ding{52}          & \ding{52}         \\
    \texttt{ietf-yang-types}               & \ding{52}         & \ding{52}          & \ding{52}         \\
    \texttt{ietf-netconf-monitoring}       & \ding{52}         & \ding{52}          & \ding{52}         \\
    \texttt{ietf-netconf-notifications}    & \ding{52}         & \ding{52}          & \ding{52}         \\
    \texttt{ietf-netconf-acm}              & \ding{52}         & \ding{52}          & \ding{52}         \\
    \texttt{ietf-netconf-with-defaults}    & \ding{52}         & \ding{52}          & \ding{52}         \\

    %\texttt{ietf-inet-types}            \cite{rfc6991:2013}   & \ding{52}         & \ding{52}         & \ding{54}         & \ding{52}         \\
    %\texttt{ietf-yang-types}            \cite{rfc6991:2013}   & \ding{52}         & \ding{52}         & \ding{54}         & \ding{52}         \\
    %\texttt{ietf-netconf-monitoring}    \cite{rfc6022:2010}   & \ding{52}         & \ding{52}         & \ding{52}         & \ding{52}         \\
    %\texttt{ietf-netconf-notifications} \cite{rfc6470:2012}   & \ding{52}         & \ding{52}         & \ding{54}         & \ding{52}         \\
    %\texttt{ietf-netconf-acm}           \cite{rfc6536:2012}   & \ding{52}         & \ding{52}         & \ding{54}         & \ding{52}         \\
    %\texttt{ietf-netconf-with-defaults} \cite{rfc6243:2011}   & \ding{52}         & \ding{52}         & \ding{52}         & \ding{52}         \\

    \bottomrule
		\end{tabular}
	\end{center}
\end{table}

Recently, the \ac{IETF} has standardized an optional underlying \ac{TLS}
transport (RFC 7589) and a NETCONF \ac{CH} mechanism (RFC 8071).  The optional
\ac{TLS} transport can be used to securely establish a NETCONF session in
situations where the mandatory \ac{SSH} transport (RFC 6242) mechanism is
unavailable.  The NETCONF \ac{CH} on the other hand is used to allow NETCONF
servers behind a \ac{NAT} to successfully establish a session with a \ac{NMS}
running a NETCONF client.  However, neither the \ac{TLS} transport nor the
\ac{CH} mechanism is widely supported by \ac{NETCONF} implementations because
the YANG data model for configuring \ac{NETCONF} clients and servers to
configure \ac{TLS} and \ac{CH} have not been standardized yet.

We recently implemented the standardized \ac{TLS} transport and \ac{CH}
mechanism in the \texttt{netopeer} and \texttt{ncclient} projects.
\texttt{netopeer} (§\ref{sec:netopeer}) is a \ac{NETCONF} client and server
implementation based on \texttt{libnetconf}~\cite{rkrejci:im:2013}.
\texttt{netopeer} is the only actively developed and maintained open-source
project implementing a \ac{NETCONF} server-side functionality.
\texttt{ncclient}~\cite{sbhushan:ipom:2009} (§\ref{sec:ncclient}) on the other
hand is a Python library that facilitates client-side scripting and
application development around the \ac{NETCONF} protocol. In this paper, we
introduce both projects. We then discuss approaches to integrate \ac{TLS}
transport (§\ref{sec:tls}) and \ac{CH} mechanism (§\ref{sec:ch}) into
\ac{NETCONF} tools and share our experiences with implementing these
mechanisms in both projects.  The implementations of the approaches described
in this paper are available online at Github. We present how implementations
of the \ac{TLS} transport and \ac{CH} mechanism allows management of home
routers (§\ref{sec:openwrt},\ref{managerouter}) using these latest additions
to the \ac{NETCONF} protocol.
%and discuss future perspectives (see~\autoref{sec:future}) on \ac{IETF}
%standardization activities that are currently being pursued in this space.

%The implementations of the approaches described in this paper are
%available~\cite{netopeer, ncclient} online.

%This effort required server-side optimizations to accommodate it to the
%limitations of the SamKnows hardware.

\section{netconf server (netopeer)} \label{sec:netopeer}

%\hl{$--$ Radek's updates on top of his IM 2013 paper.}

\begin{figure}[t!]
  \centering
  \includegraphics* [width=.7\linewidth]{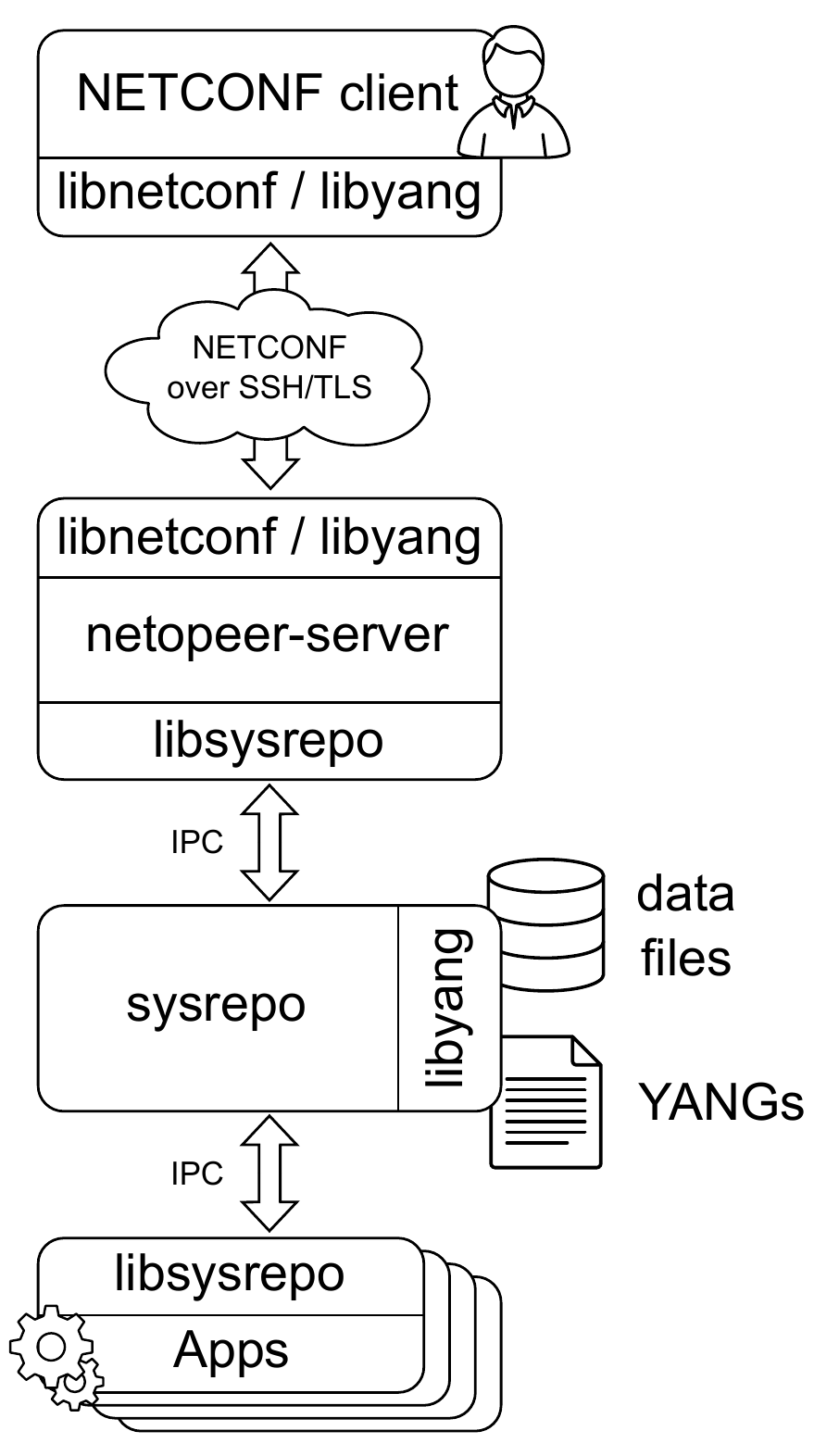}

  \caption{\textbf{netopeer building blocks and their interconnections.
  libnetconf implements the NETCONF protocol specification, while libyang
  serves as a parser and validator of YANG schemas. sysrepo supports
  applications to store and manipulate configuration and operational
  data modeled by YANG. netopeer-server is the NETCONF server implementation
  that integrates libnetconf and sysrepo.}}

  \label{fig:netopeer-arch}
\end{figure}

The \texttt{netopeer} project provides a \ac{NETCONF} server implementation
that allows application developers to integrate remote configuration
functionality using the \ac{NETCONF} protocol within their products. The
project has evolved over the years and there are multiple building blocks that
can be used independently as shown in Fig.~\ref{fig:netopeer-arch}. While each
building block is written in C, there are bindings available for various
languages to allow applications written in different programming languages to
use the offered functionality. We next describe each building block of the
\texttt{netopeer} project.

\subsection{libyang}

%Support for default values in the instance data (RFC 6243).
%Support for YANG extensions.
%Support for YANG Metadata (RFC 7952).

\texttt{libyang} is YANG data modeling language parser and toolkit. A need for
better handling of the YANG 1.0 (RFC 6020) and YANG 1.1 (RFC 7950) schemas and
manipulation of instance data in the \texttt{netopeer} server was felt which
led to the creation of the \texttt{libyang} toolkit.  While the main
functionality is to provide access and allow manipulation with the data
modeled by YANG, \texttt{libyang} also has the capability to parse and
validate YANG (and YIN) schemas. It can also parse, validate and print
instance data in XML and JSON (RFC 7951) format. \texttt{libyang} includes
\texttt{yanglint}, which is a feature rich tool for validation and conversion
of the schemas and YANG modeled data. \texttt{libyang} supports default values
in the instance data (RFC 6243) and YANG metadata (RFC 7952) and extensions.

\subsection{libnetconf}

\texttt{libnetconf} is a library that implements the \ac{NETCONF} protocol. It
can be used by \ac{NETCONF} clients as well as servers to create, send,
receive, process and respond to \ac{NETCONF} requests as well as asynchronous
event notifications. \texttt{libnetconf} is \ac{NETCONF} v1.0 (RFC 4741) and
v1.1 (RFC 6241) compliant thereby providing support for several basic
\ac{NETCONF} capabilities such as writable-running, candidate configuration,
validate capability, distinct startup and URL capability as defined in RFC
6241. \texttt{libnetconf} supports the mandatory underlying \ac{SSH} (RFC
6242) transport mechanism and support for DNSSEC SSH key fingerprints (RFC
4255). Further features include support for NETCONF Event Notifications (RFC
5277, RFC 6470), with-defaults capability (RFC 6243) and NETCONF access
control (RFC 6536). \texttt{libnetconf} also provides bindings for
applications written in the Python programming language. We further added to
this feature set, by implemnenting the optional NETCONF over
TLS (RFC 7589) transport mechanism and NETCONF \ac{CH} mechanism using
reverse \ac{TLS} and \ac{SSH} (RFC 8071).

Based on experiences with \texttt{libnetconf}, the next generation of this
library (\texttt{libnetconf2}) based on \texttt{libyang} is currently under
development. Originally, \texttt{libnetconf} used \texttt{libxml2} to
represent \ac{NETCONF} data. Although the \ac{NETCONF} (YANG modeled) data can
be represented as XML data, it is limited when compared with a generic XML
data format where mixed content is not allowed. As such, it is inefficient to
parse and store this data in generic XML structures. The way of storing and
manipulating the data has been changed in \texttt{libnetconf2} which uses the
\texttt{libyang} toolkit.  With \texttt{libnetconf2}, the implementation of
the \ac{NETCONF} datastores as well as the mechanism used to connect the
datastore with applications (TransAPI) have been moved out of the library into
the \texttt{sysrepo} project described below.

\subsection{sysrepo}

%Applications can currently use C language API of sysrepo Client Library to
%access the configuration in the datastore, but the support for other
%programming languages is planed for later too (since sysrepo uses Google
%Protocol Buffers as the interface between the datastore and client library,
%writing of a native client library for any programing language that supports
%GPB is possible).  Sysrepo can be easily integrated with management agents
%such as NETCONF or RESTCONF servers, using the same client library API that
%applications use to access their configuration. As of now, sysrepo is
%integrated with the Netopeer 2 NETCONF server. This means that applications
%that use sysrepo to store their configuration can automatically benefit from
%the ability to being controlled via NETCONF.

\texttt{sysrepo} is a YANG-based configuration and operational state data
store for GNU/Linux applications.  The main purpose is to support applications
with simple and unified access to their configuration data.  As such,
applications (instead of using flat configuration files) can use
\texttt{sysrepo} to store their configuration modeled by the provided YANG
model. \texttt{sysrepo} provides access to administrators to manipulate with
the configuration data of the connected applications. Using YANG schemas,
\texttt{sysrepo} validates configuration data and all the performed changes.
In this way, it ensures consistency of the data stored in the datastore and
enforce data constraints defined by the YANG models. It interconnects
administrators (including automated control systems) with the applications by
announcing configuration changes to the applications and providing event
notifications and status data from applications to administrators.

\texttt{sysrepo} implements the \ac{NETCONF} datastore previously present in
\texttt{libnetconf}. In contrast to previous implementation, it is based on
\texttt{libyang} and able to run as a standalone local daemon. However, such a
local \texttt{sysrepo} instance can be extended by \texttt{netopeer-server}
providing remote access to the datastore via \ac{NETCONF} protocol.

\subsection{netopeer-server}

\texttt{netopeer-server} is the \ac{NETCONF} server implementation. It
integrates the \texttt{sysrepo} datastore and covers \texttt{libnetconf}
functionality to interconnect datastores with \ac{NETCONF} clients.  Using
\texttt{libnetconf} functions, \texttt{netopeer-server} integrates \ac{SSH}
and \ac{TLS} servers listening for incoming \ac{NETCONF} connections. The
server behavior is controlled through implementation of several standard
\ac{IETF} YANG schemas covering \ac{NETCONF} server configuration. As such,
the \texttt{netopeer-server} behaves as controlling as well as a controlled
application when comminucating with \texttt{sysrepo}. The schemas cover even
the \ac{CH} mechanism. It enables \texttt{netopeer-server} not only listen for
incoming connections but also actively open and maintain connections to
\ac{NMS} according to the specified configuration.

%Similarly as libnetconf, the netopeer-server also evolved and after complete
%redesign it is now named netopeer2-server.

\section{netconf client (ncclient)} \label{sec:ncclient}

\texttt{ncclient}~\cite{sbhushan:ipom:2009}  is a Python library that
facilitates client-side scripting and application development around the
\ac{NETCONF} protocol.  \texttt{ncclient} is an open-source project hosted on
github with an Apache 2.0 licence. Bhushan \emph{et al.}
in~\cite{sbhushan:ipom:2009} describe the design and first prototype
implementation of \texttt{ncclient}. The project has significantly evolved
over several years of development effort and adoption in the industry. For
instance, initially, \texttt{ncclient} provided support for \ac{NETCONF} v1.0
(RFC 4741). As network equipment vendors adopted and implemented the
\ac{NETCONF} protocol, \texttt{ncclient} started to gain visibility and
attract interest. However, minor deviations in the implementation of
\ac{NETCONF} v1.0 on behalf of the vendors made each of them fork the project
and implement their own flavor of \texttt{ncclient}. A major release merged
the vendors' changes and the idea of modular device handlers was introduced.
Each vendor can define their individual handling for \texttt{ssh} connection
parameters, device operations, reply parsing, and proprietary server
capabilities. Later releases added compatibility with Python 3 and NETCONF
event notifications (RFC 5277). A milestone in the development history of
\texttt{ncclient} is our implementation of NETCONF v1.1 (RFC 6241) which adds
support for handling chunked frames (RFC 6242).  As \texttt{ncclient} was
already established, we made sure that backwards compatibility is maintained
to allow clients to fallback to NETCONF v1.0 end of message framing scheme
(RFC 4742) so as to provide support for server implementations that have not
yet adopted and advertise the NETCONF v1.1 capability.

% Initial versions of \texttt{ncclient} assumed that the \ac{NETCONF} server
% is configured as a subsystem in the \texttt{ssh} server implementation.
% Contributing to \texttt{ncclient's} evolution, we introduced remote command
% invocation as a fallback when \ac{NETCONF} subsystem invocation fails.

\texttt{ncclient} device handler provides customized solutions for
Alcatel-Lucent/Nokia 7x50, Cisco CSR, Cisco IOS-XE, Cisco IOS-XR, Cisco Nexus,
Huawei, HP Comware and Juniper devices. \texttt{ncclient} is the core
%component of Juniper's network automation framework Junos PyEZ~\cite{pyez},
component of Juniper's network automation framework Junos PyEZ, the core
component of Openstack's networking as a service project
%Neutron~\cite{ostkneutron}, the core component of the open-source network
Neutron, the core component of the open-source network virtualization platform
%OpenContrail~\cite{opencontrail}, the core component of GRNET's and GEANT's
OpenContrail, the core component of GRNET's and GEANT's Firewall on Demand
flowspec-based \ac{DDoS} mitigation
%services~\cite{fodgrnet}~\cite{Poulopoulos:Mamalis:Polyrakis:FireCircle:12}~\cite{fodgeant}
services~\cite{Poulopoulos:Mamalis:Polyrakis:FireCircle:12}
%and the core component of Ansible's NETCONF role~\cite{ansiblenetconf}. It is
and the core component of Ansible's \ac{NETCONF} role. It is also used in
several academic and corporate network management projects.

\texttt{ncclient} is currently one of the most comprehensive \ac{NETCONF}
client implementations not only in the open source community but also in the
industry. From a development standpoint, it is actively developed and
maintained with new features and performance improvements. What makes the case
of \texttt{ncclient} unique is the combined contribution of industry, academia
and private individuals.

\section{netconf over tls transport} \label{sec:tls}
%\section{\ac{NETCONF} over \ac{TLS} Transport}

%\todovb{some other limitations/restrictions of the spec?}

%In this approach, the \ac{NETCONF} developer perspective, it is not
%neccessary to know anything about a \ac{TLS} communication in this case.

\begin{figure}[t!]
  \begin{center}
     \includegraphics* [width=1.0\linewidth]{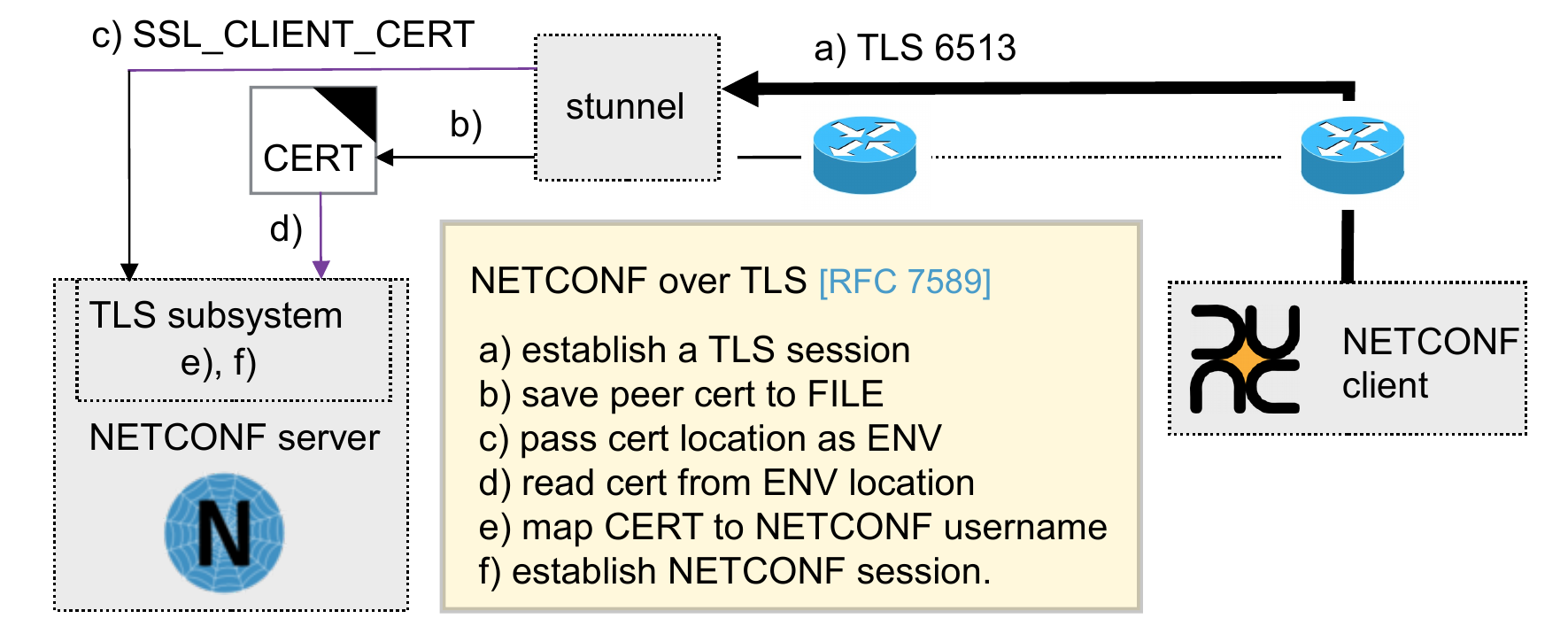}
  \end{center}

  \caption{\textbf{ncclient establishing a \ac{NETCONF} over \ac{TLS} session
  with a \ac{NETCONF} server using stunnel as a \ac{TLS} proxy server.}}

  \label{fig:nc-tls-stunnel}
\end{figure}

We describe the integration of \ac{TLS} transport mechanism in both
\texttt{netopeer} and \texttt{ncclient} projects.  There are two general
approaches to integrate \ac{TLS} support within a \ac{NETCONF} server. The
first approach is to use a standalone \ac{TLS} proxy prepended before the
\ac{NETCONF} server as shown in Fig.~\ref{fig:nc-tls-stunnel}.  We use
\texttt{stunnel} since the \ac{NETCONF} over \ac{TLS} transport specification
(RFC 7589) mandates the use of at least \ac{TLS} v1.2 with
mutual-authentication to secure the session. The \ac{TLS} server
(\texttt{stunnel}) works as a proxy, which encrypts (and decrypts) the
communication between the \ac{NETCONF} client and the server.  However,
client-authentication depends on the content of the \ac{NETCONF}
configuration. As such, it is necessary to either implement the \ac{NETCONF}
server configuration data directly inside the \ac{TLS} proxy
(\texttt{stunnel}) or pass the client certificate back to the \ac{NETCONF}
server (\texttt{netopeer}) during the \ac{TLS} handshake. The former approach
requires larger modifications of the \ac{TLS} proxy, which is the reason we
initially implemented the latter approach within \texttt{netopeer}. In this
way, when \texttt{ncclient} initiates a \ac{TLS} session (see
Fig.~\ref{fig:nc-tls-stunnel}), the client certificate is checked within
\texttt{stunnel} and then passed into the \texttt{netopeer} server to map the
certificate to a specific user and authorise the client for further
communication.

\begin{figure}[t!]
  \begin{center}
     \includegraphics* [width=1.0\linewidth]{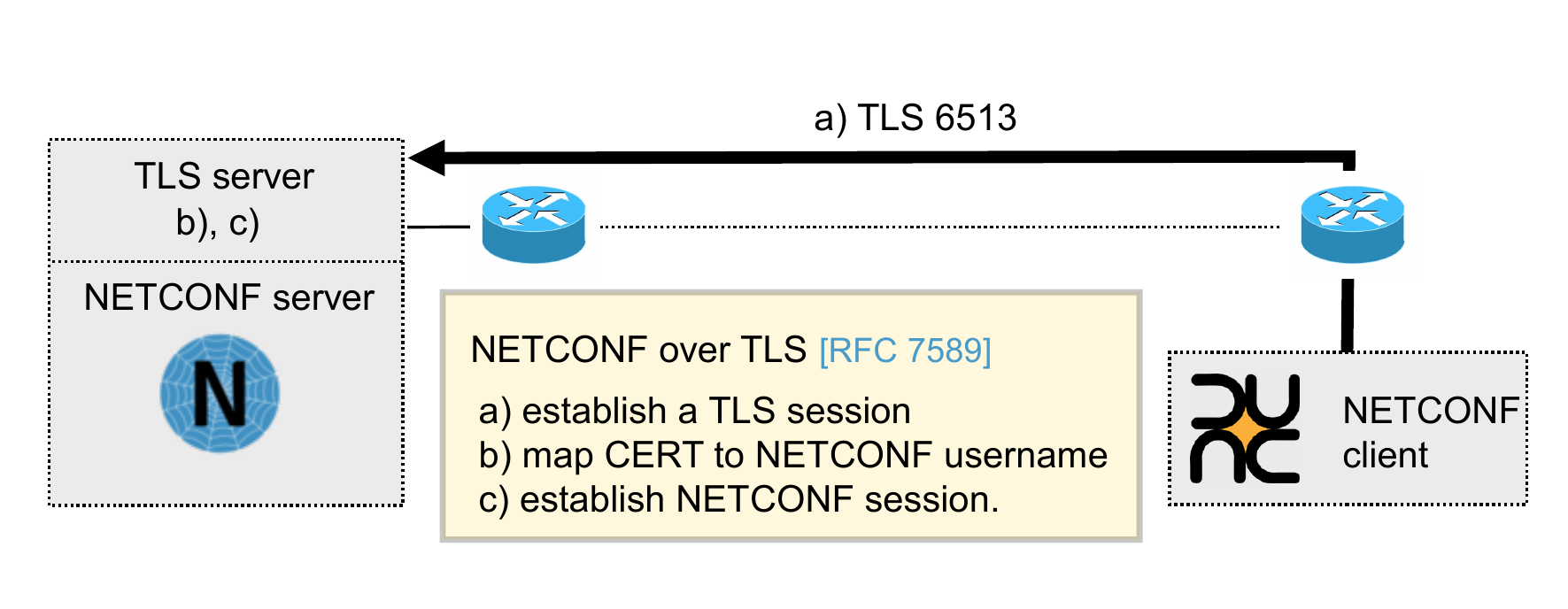}
  \end{center}

  \caption{\textbf{ncclient establishing a \ac{NETCONF} over \ac{TLS} session with a
  NETCONF server using an integrated TLS server.}}

  \label{fig:nc-tls-integrated}
\end{figure}

The second approach is to integrate \ac{TLS} support on the server-side by
implementing a \ac{TLS} server directly within the \ac{NETCONF} server
process by using a \ac{TLS} library as shown in Fig.~\ref{fig:nc-tls-integrated}.
This approach significantly simplifies the
\ac{NETCONF} server-side but requires additional effort to correctly implement
a \ac{TLS} server. It is still necessary to authorize the connecting client
according to the current \ac{NETCONF} server configuration. However, the
certificate processing is done directly in the \ac{NETCONF} server process.
This makes it simpler to access the server configuration data to map the
certificate to the user and authenticate the client.  In the \texttt{netopeer}
server, we first implemented \ac{TLS} support using the \texttt{stunnel}
standalone \ac{TLS} proxy server. However, to simplify the server design and
the deployment process, \ac{TLS} server was later moved directly into the
\texttt{netopeer} server by using \texttt{openssl} toolkit.

The client-side (\texttt{ncclient}) update was simpler with an additional
\ac{TLS} transport mechanism that largely offloaded the \ac{TLS} support to
the Python \texttt{ssl} library with strict authentication support. In this
way, \texttt{ncclient} can now establish a secure NETCONF session with
\texttt{netopeer} over \ac{TLS}.

%\subsection{NETCONF client}

%\texttt{ncclient}

%\begin{enumerate}
  %\item A new TLS transport mechanism.
  %\item Authenticate the server certificate.
%\end{enumerate}

\section{netconf call home} \label{sec:ch}

%The YANG data model is currently a work in progress within the \ac{NETCONF} WG
%at the \ac{IETF}.

\begin{figure}[t!]
  \begin{center}
     \includegraphics* [width=1.0\linewidth]{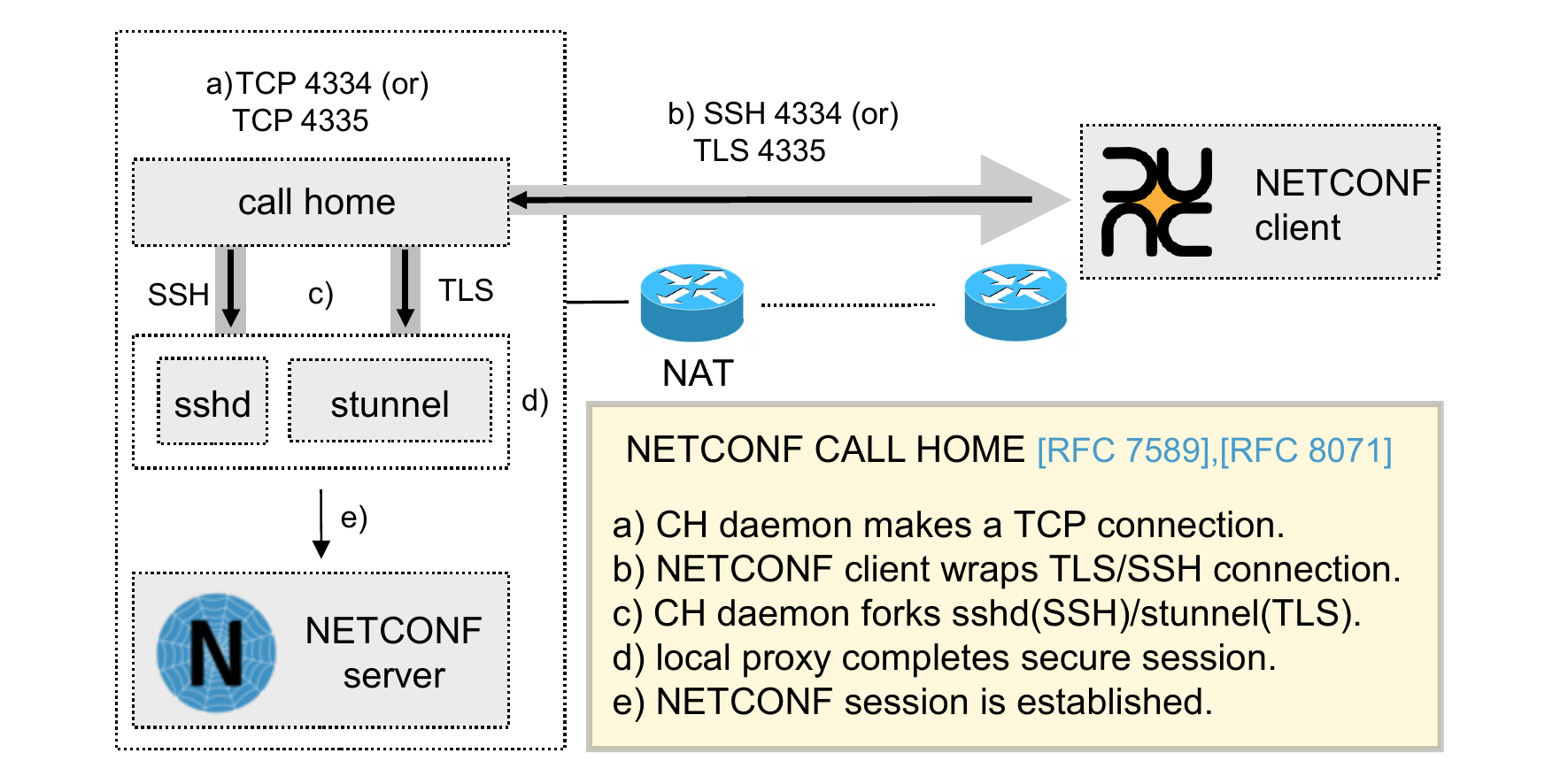}
  \end{center}

  \caption{\textbf{\ac{NETCONF} \ac{CH} using a standalone \ac{CH} daemon.}}
%   The call home daemon establishing a \ac{TCP} connection to
%   ncclient, which uses it to reverse roles and establish a TLS (or SSH)
%   connection back to the initiator.  The call home daemon on receiving the
%   request forks stunnel or the ssh daemon depending on the type of incoming
%   connection. The local proxy completes the secure connection with ncclient and
%   spawns the NETCONF subsystem to establish a NETCONF session.}

  \label{fig:nc-ch-proxy}
\end{figure}

\begin{figure}[t!]
  \begin{center}
     \includegraphics* [width=1.0\linewidth]{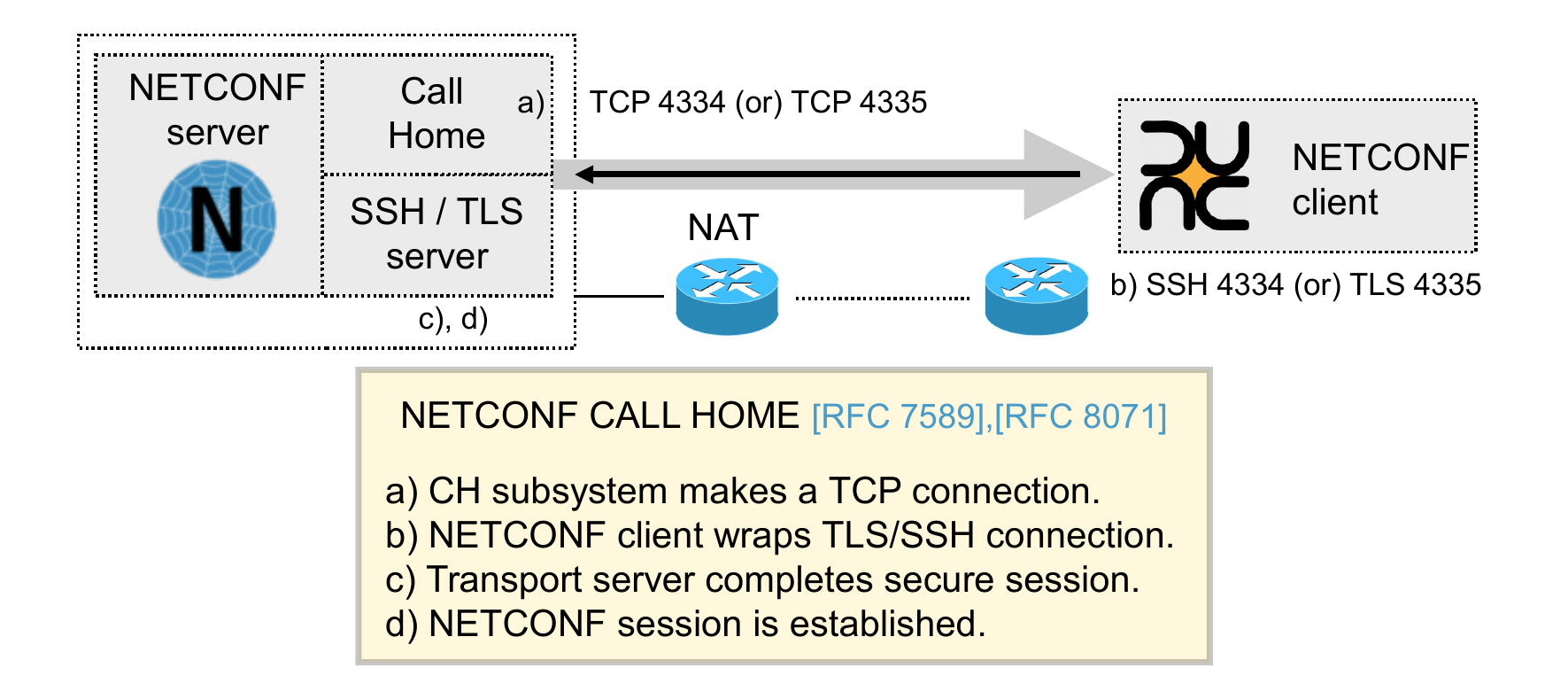}
  \end{center}

  \caption{\textbf{\ac{NETCONF} \ac{CH} using an integrated \ac{CH} daemon and
  transport servers within a \ac{NETCONF} server.}}

  \label{fig:nc-ch}
\end{figure}

The \ac{CH} mechanism (RFC 8071) allows a managed device deployed behind a
\ac{NAT} running a \ac{NETCONF} server to successfully establish a session
with a \ac{NMS} running a \ac{NETCONF} client.  The \ac{CH} mechanism can be
secured either over \ac{SSH} or \ac{TLS} transport. The \ac{CH} mechanism
requires implementation of a \ac{CH} daemon on the server-side
(\texttt{netopeer}) and a TCP listen mode on the client-side
(\texttt{ncclient}). In this way, the underlying TCP connection is initiated
by the server thereby punching a hole in the \ac{NAT} gateway. Once the TCP
connection is established, the roles are reversed and the control is delegated
to the \ac{NETCONF} client which initiates a \ac{TLS} (or a \ac{SSH}) secure
session and eventually the \ac{NETCONF} session on top of the established
secure transport. The \ac{NETCONF} server also needs to support the server
configuration YANG data model (work in progress) that specifies how the
\ac{CH} connections are maintained. This allows the \ac{NETCONF} server to
monitor call home sessions and re-establish the connection in a standardized
way. As in case of the \ac{TLS} transport implementation, \ac{CH} can be
implemented as a standalone process or it can be fully integrated into the
\ac{NETCONF} server process. The standalone \ac{CH} daemon on receiving the
request forks a transport server process (\texttt{sshd} or \texttt{stunnel})
depending on the type of the incoming connection as shown in
Fig.~\ref{fig:nc-ch-proxy}. Alternatively, the integrated \ac{CH} daemon can
pass the connection data directly into the \ac{NETCONF} server with the
integrated transport server as shown in Fig.~\ref{fig:nc-ch}.  The pros and
cons of both approaches are similar to the case of integrating the \ac{TLS}
transport.

On the client-side, a \texttt{--callhome} switch is implemented to start
\texttt{ncclient} in TCP listen mode.  \texttt{ncclient} listens on TCP port
4334 (\ac{SSH} transport) and on TCP port 4335 (\ac{TLS} transport)
simultaneously.  A \ac{CH} from \texttt{netopeer} establishes the TCP
connection. The roles are reversed at this stage whereby \texttt{ncclient}
passes on the connected socket and depending on the port used,
\texttt{ncclient} initiates a \ac{TLS} (or \ac{SSH}) session and a
\ac{NETCONF} session on top of the secure transport.

%While the figures show combination of each
%implementation approach with the same approach used for the transport servers,
%it can be arbitrarily combined.

%\subsection{NETCONF client}
%\todovb{modifications to the client}
% \section{Limitations}
% I believe that limitations and pros/cons are described inline

\section{Deployment on a Home Router} \label{sec:openwrt}
% 
% \begin{figure}[t!]
%   \centering
%  \includegraphics* [width=1.0\linewidth]{nc-samknows}
% 
%   \caption{\textbf{ncclient establishing a \ac{NETCONF} session with a
%   \texttt{netopeer} NETCONF server running on a TP-Link device. This setup can
%   now be used to also manage the home router from the outside world.}}
% 
%   \label{fig:nc-samknows}
% \end{figure}

\ac{NETCONF} (RFC 6241) can function as a control protocol to manage home
gateways. In order to understand the memory impact and possibility of using
\ac{NETCONF} on these low-end consumer devices, we deployed a \ac{NETCONF}
server on an TP-Link device running OpenWrt.
% as shown in Fig.~\ref{fig:nc-samknows}.
We describe a number of technical challenges we
faced during this process. These real-world issues led to improvements of both
\texttt{ncclient}, \texttt{netopeer} and related OpenWrt packages.

%Furthermore, it is written in C so it can be easily
%optimized for limited SamKnows hardware.

\subsection{Optimizations}

%\subsection{C library restrictions}

The \texttt{libnetconf} library (part of the \texttt{netopeer} project) uses
GNU extensions and some other features (such as \texttt{utmpx}) that are not
available in \texttt{uClibc}, the C library available on the OpenWrt platform.
Therefore, \texttt{libnetconf} was modified to detect availability of such
extensions. In case of missing C library features, \texttt{libnetconf} was
enhanced with alternative ways to achieve the requested functionality.

%\subsection{SSH package misconfigurations}

There is a misconfiguration in the \texttt{libssh2} OpenWrt package. The
package is built with static and dynamic library support by default. However,
the package although does export dynamic libaries, it does not export
static libraries into staging directories. This makes the static library
unusable by default. We updated the package sources and informed the OpenWrt
package maintainers to get this resolved upstream.

%\subsection{NETCONF event notifications}

\texttt{libnetconf} supports \ac{NETCONF} event notifications (RFC 5277).  It
uses \texttt{dbus} to share information about these events between
\texttt{libnetconf} instances. However, not all versions of OpenWrt support
\texttt{dbus}. Besides modifying the \texttt{libnetconf} build process to
switch off support for \ac{NETCONF} notifications, the \texttt{libnetconf}
event notifications codebase was also simplified and \texttt{dbus} is no more
required.

%\subsection{Optimizations}
% applies to quite old netopeer-server version and confuses reader here with
% all that information about netopeer modifications since that time

The \ac{NETCONF} server binary before being deployed was 2.6Mib in size.
A considerable part of this size was occupied by the statically linked code
not necessary for the server side applications. By optimizing the building
process and providing alternative implementation of several features in
\texttt{libnetconf}, we were able to reduce the server binary to 1.5MiB in size.

% The \ac{NETCONF} server binary before being deployed was 2.6Mib in size.  A
% considerable part of this size was occupied by the statically linked
% \texttt{libssh2} library. \texttt{libssh2}, however, is used only by
% client-side functions to establish an \texttt{ssh} connection to the server.
% For server-side applications, this functionality is needless. Furthermore, on
% client-side this requirement can be alternated by utilizing a standalone
% \texttt{ssh} client application. Therefore, we updated the \texttt{libnetconf}
% build process with a \texttt{-–disable-libssh2} option that dramatically
% reduces the binary size of the server implementation.  Eventually, the
% \texttt{libnetconf} server depends on \texttt{libevent} and \texttt{libxml2}.
% We statically linked these libraries into the final binary because they are
% not available on OpenWrt devices. We stripped the final binary to remove any
% symbol table information and debugging symbols. The server reduced to 1.5MiB
% in size.

%\subsection{SSH subsystems}

When the \ac{NETCONF} server uses standalone transport servers instead of their
integration, it is necessary to hook it as a subsystem in the \texttt{ssh} server
implementation. However,
OpenWrt deploys \texttt{dropbear} as a \texttt{ssh} server implementation.
\texttt{dropbear} is a tiny \texttt{ssh} server implementation that is mostly
suited for embedded Linux, such as wireless routers, but they do not have any
support to hook subsystems. OpenWrt provides packages for alternative OpenSSH
server implementation, however there is not enough flash memory space
available to run a OpenSSH server on home routers. In order to circumvent the
issue, we deployed a separate \texttt{dropbear} instance to handle
\ac{NETCONF} specific calls on port 830. Anyway, the issue was later overcome
by integrating the trnsport servers directly into the \texttt{netopeer-server}.

% A \ac{NETCONF} server must be hooked as a subsystem in the \texttt{ssh} server
% implementation. This allows the \texttt{ssh} server to delegate all
% \ac{NETCONF} calls to the server daemon on standard port 830.  However,
% OpenWrt deploys \texttt{dropbear} as a \texttt{ssh} server implementation.
% \texttt{dropbear} is a tiny \texttt{ssh} server implementation that is mostly
% suited for embedded Linux, such as wireless routers, but they do not have any
% support to hook subsystems. OpenWrt provides packages for alternative OpenSSH
% server implementation, however there is not enough flash memory space
% available to run a OpenSSH server on home routers. In order to circumvent the
% issue, we deployed a separate \texttt{dropbear} instance to handle
% \ac{NETCONF} specific calls on port 830.  Remote commands, such as launching
% the \ac{NETCONF} server daemon can be issued on this \texttt{ssh} channel,
% which allows one to interact with the server daemon from outside.  However,
% the client does not know where the server is located on the remote machine. As
% a workaround, we created a symbolic link from the shell's search PATH to the
% custom remote location of the server location, so that the server daemon is
% executed when the client drops down the \texttt{ash} shell.

\section{Home Router Management}
\label{managerouter}

We use the core system data model (RFC 7317) standardized by the \ac{IETF} to
configure such home gateways. This YANG data model specifies configuration and
state data used to set and retrieve information about the system
identification, time and users management as well as a DNS resolver.
Furthermore, the model defines new \ac{NETCONF} operations for a device
restart and shutdown.  We prepared a \texttt{libnetconf} core system
\texttt{transAPI} module for our \ac{NETCONF} server. The module is a set of
callbacks employed when specific parts of a configuration data are changed.
For example, when a user changes timezone settings in the \texttt{running}
datastore, a specific callback function applying this parameter to the home
gateway is automatically called. Similarly for \ac{NETCONF} RPC operations,
for example, when the system-restart operation is requested by a client,
\texttt{libnetconf} passes the program control to the appropriate
\texttt{transAPI} module function that performs a reboot of the home gateway.

The latest Netopeer works very similar way but the datastore and the interaction
with applications (core system module) is moved into sysrepo. It was rapidly
improved to e.g. provide access to the configuration data even if the \ac{NETCONF}
server or even sysrepo daemon are not running. The API is richer and more granular
to cover more use cases.

\section{Conclusion and Future Perspectives} \label{sec:conclusion}

In this article, we presented an overview of the NETCONF and YANG feature
support in popular commercial and open-source implementations. We showed that
\texttt{netopeer} (a NETCONF server) and \texttt{ncclient} (a NETCONF client)
open-source projects have evolved over several years of development effort. We
captured the significant developments made within these projects.  We further
described how we implemented the recently standardized \ac{NETCONF} \ac{TLS}
transport~\cite{rfc7589} and \ac{CH} mechanisms~\cite{rfc8071} within both
these projects. Towards the end, we demonstrated how these features allow home
routers to be managed using the \ac{NETCONF} protocol.

We envision a future where it would be possible to implement measurement
capabilities directly inside home routers circumventing the need to deploy
separate measurement devices~\cite{bajpai:comst:2015, bajpai:ccr:2015}.  The
management of such home routers will thus involve configuration and scheduling
of measurements using a control protocol.  The IETF has standardized a YANG
data model~\cite{rfc8194} that can be used to configure and schedule
measurements on home routers. RESTCONF~\cite{RFC8040} could be the control
protocol that can be used with the YANG data model to configure home routers
and report measurement results using event notifications.  RESTCONF uses HTTP
methods to provide operations and capabilities (similar to NETCONF) on a
datastore specified by the YANG schema.  We expect our implementation effort
and demonstration of possibility of using such protocols to manage home
routers serves as input towards progressing standards development within the
\ac{IETF}.

%and provided perspectives on future standards work that is currently being
%pursued in this space within the \ac{IETF}.

%The \ac{IETF} \ac{LMAP} working group is standardizing a framework for
%large-scale measurement platforms.  Such a framework will make

%for managing these devices. 
%has been approved by the IESG, while work is currently
%underway~\cite{draft-ietf-lmap-restconf} that describes how

%The \texttt{ncclient} and \texttt{netopeer} source code
%is available online.

%The \texttt{ncclient}~\cite{ncclient} and \texttt{netopeer}~\cite{netopeer}
%source code is available online.

\section*{Acknowledgements}
\label{sec:acknowledgements}

We like to thank J\"{u}rgen Sch\"{o}nw\"{a}lder for his feedback. Martin
Björklund (Tail-f) and Andy Bierman (YumaWorks) for providing us with licensed
copies of their commercial \ac{NETCONF} implementations.
%----------------------------------------------------------------------

\bibliographystyle{IEEEtran}
\bibliography{index}

\vfill

% that's all folks
\end{document}